\documentclass[aps,pre,float,twocolumn]{revtex4-1}

\usepackage{amsmath,bm,epsfig}

\def\p{\partial}

\def\a{\alpha}\def\k{{\bm k}}
\def\b{\beta}

\def\o{\omega}

\def\be{\begin{equation}}
\def\BE{\begin{equation}}
\def\ba{\begin{array}}
\def\ee{\end{equation}}
\def\EE{\end{equation}}
\def\ea{\end{array}}
\def\bea {\begin{eqnarray}}
\def\eea {\end{eqnarray}}
\def\BEA {\begin{eqnarray}}
\def\EEA {\end{eqnarray}}
\def\bean{\begin{eqnarray*}}
\def\eean{\end{eqnarray*}}

\def\Fbox#1{\vskip1ex\hbox to 8.5cm{\hfil\fboxsep0.3cm\fbox{%
  \parbox{8.0cm}{#1}}\hfil}\vskip1ex\noindent}  

\let\*\cdot
\def\<{\left\langle} \def\>{\right\rangle} \def\({\left(} \def\){\right)}
\let\p\partial \let\~\widetilde \let\^\widehat 

\newcommand{\B}[1]{{\bm{#1}}}
\newcommand{\C}[1]{{\mathcal{#1}}}    

\def\be{\begin{equation}}\def\ee{\end{equation}}
\def\bea{\begin{eqnarray}}\def\eea{\end{eqnarray}}
\def\bse{\begin{subequations}}\def\ese{\end{subequations}}
\newcommand{\BSE}[1]{\begin{subequations}\label{#1}}
\def\bc{\begin{cases}}\def\ec{\end{cases}}

\newcommand{\Eq}[1]{Eq.~(\ref{#1})}
\newcommand{\Eqs}[1]{Eqs.~(\ref{#1})}

\renewcommand{\sb}[1]{_{\rm{#1}}}  

\newcommand{\eq}[1]{(\ref{#1})}

\usepackage{amssymb,amsmath}

\def\BSE{\begin{subequations}}\def\ESE{\end{subequations}}
\let \= \equiv

\def\p{\partial}

\def\a{\alpha}
\def\b{\beta}

\def\o{\omega}

\def\be{\begin{equation}}       \def\ba{\begin{array}}

\def\ee{\end{equation}}         \def\ea{\end{array}}

\def\bea {\begin{eqnarray}}      \def\eea {\end{eqnarray}}

\def\bean{\begin{eqnarray*}}    \def\eean{\end{eqnarray*}}

\def\la  {\lambda}

\def\<{\langle} \def\({\left(}  \def\>{\rangle} \def\){\right)}

\newtheorem{exi}{Example}

\begin{document}

\title{Discrete and mesoscopic regimes of finite-size   wave turbulence}
\author{ V. S. L'vov$^{\dag}$ and  S. Nazarenko$^{\ddag}$}

  \affiliation{
 $^\dag$ Department
of Chemical Physics, The Weizmann Institute of Science, Rehovot
76100, Israel \\
$^\ddag$Mathematics Institute, University of Warwick, Coventry CV4-7AL,
UK}

\begin{abstract}
Bounding volume results in discreteness of eigenmodes in wave systems. This leads to a depletion or
complete loss of wave resonances (three-wave, four-wave, etc.), which has a strong effect on \em Wave Turbulence, \em (WT) i.e. on the statistical behavior of broadband sets of weakly nonlinear waves. This paper describes three different regimes of WT realizable for different levels of the wave excitations:
\emph{Discrete, mesoscopic and kinetic WT}.  \emph{Discrete WT}  comprises  chaotic dynamics of interacting wave ``clusters"  consisting of discrete (often finite)
    number of connected resonant wave triads (or quarters).
    \emph{Kinetic WT} refers to the infinite-box theory, described by well-known wave-kinetic equations.
     \emph{Mesoscopic WT} is a regime in which either the discrete and the kinetic evolutions alternate, or when none of these two
      types is purely realized.
      We argue that in  mesoscopic systems the wave spectrum experiences a \em sandpile \em behavior.
      Importantly, the mesoscopic regime is realized for a \em broad range \em of wave amplitudes which typically
      spans over several orders on magnitude, and not just for a particular intermediate level.

\end{abstract}


\maketitle

\section*{\label{s:intro} Introduction}

Dispersive waves play a crucial role   in a vast range of physical
applications, from quantum to classical systems, from microscopic to astrophysical scales.
For example, Kelvin waves propagating on quantized vortex lines
provide an essential mechanism of turbulent energy cascades in
quantum turbulence in cryogenic Helium
\cite{VinenNiemela,kozik_svistunov,Vin03-PRL,Naz_kelvin,llnr09,ln_spectrum};
water waves aid
the momentum and energy transfers from wind to ocean \cite{janssen_book};
internal waves on density
stratifications and inertial waves due to rotation are important in
turbulence behavior and mixing in planetary atmospheres and oceans
\cite{Zeitlin,yura,galtier_inertial}; planetary Rossby waves are important for
the weather and climate evolutions \cite{rossby}; and Alfven waves
are ubiquitous in turbulence of solar wind and interstellar medium
\cite{Iroshnikov,mhd_book,ng,galtier_mhd,goldreich,goldreich_s}.
More often than not, nonlinear interaction of different wave modes
is important in these and other applications, and there has been a
significant amount of work done in the past to describe evolution of
such interacting wave systems. If the number of excited modes is large,
they experience random evolutions, which must be described by a statistical
theory.
\emph{Weak Wave Turbulence} refers to such a statistical
theory for weakly nonlinear dispersive waves in unbounded domains
\cite{ZLF}. This approach was initiated by Peierls in 1929 to
describe phonons in anharmonic crystals \cite{peierls},
 and it was reinvigorated in
1960's in plasma physics
\cite{galeev_sagdeev,vedenov,zaslavski_sagdeev} and in
the theory of water waves \cite{zakh_fil,fil}. By now,
it has been applied to description of a great variety of
physical phenomena, from synoptic Rossby waves
\cite{longuet_higgens,zakharov_piterbarg,monin_piterbarg,balk_nazarenko}
to magneto-hydrodynamic turbulence \cite{galtier_mhd,ng,goldreich}, to
acoustic waves \cite{ZakharovSagdeev},  to waves
in stratified \cite{Zeitlin,yura} and rotating fluids
\cite{galtier_inertial}, and many other physical wave systems.

On the other hand, it has become increasingly
   clear that in the majority laboratory experiments and numerical simulations of nonlinear dispersive wave
    systems the discreteness of the wave-number space due to a finite size is a crucially  important factor which
     causes the system behave differently from the predictions of the classical theory of wave turbulence
      based on the continuous (infinite domain) limit
      \cite{denis,fauve,NazarLvov2006,Tan2004,conn,lidya,Naz-2006,z2,slavedMHD}.
      Moreover, similar behavior often occurs in nature when waves
      are bounded, e.g. for planetary Rossby waves bounded by the finite planet radius \cite{KPR}.

      Description of transition from regular to random regimes and characterization of the intermediate states
      where both regular and random wave motions are present and mutually inter-connected, is an intriguing and
      challenging problem.  Such intermediate states where the number of waves is big and yet the discreteness
      of the wavenumber space still remains important are called   \emph{Discrete} and  \emph{Mesoscopic Wave Turbulence}.


\section{\label{ss:HF} Weakly interacting waves}

\subsection{\label{ss:NM} Normal modes of linearized problem}

An evolution equation
  is called
\emph{dispersive} if its linear part has wave-like solutions
$\psi(\bm r,t)$, that depend on the coordinate in the $d$-dimensional
physical space, $\bm r \in \mathbb{R}^d$ and time $t$ as follows:%
 \begin{subequations} \label{NM} \BEA \label{PW} \psi ({\bm
r},t)&=& A_\k e^{i( {\bm k}\cdot{ \bm  r} - \o\, t)} + \mbox{c.c.}\\
\nonumber  &=& |A_\k| \cos({\bm k}\cdot{ \bm  r} - \o\, t+\varphi)\,,
\eea
where  ``c.c."  means ``complex conjugate".
Here  $A_\k=|A_\k| \exp (i\varphi) \in \mathbb{C}$ is a constant wave
 amplitude, $\varphi$ is the wave phase,
 ${ \bm  k} \in \mathbb{R}^d$ is
a   wavevector  and wave
frequency $\o \equiv \o({ \bm  k}) \in \mathbb{R}$ is such that
$
 \big |  {\p^2 \o}\big / {\p {k_i} \p {k_j}} \big | \not\equiv 0 $,
 where $k_i$ and $
k_j$ are components of ${ \bm  k}$. Physically, the
latter condition means that wave packets with different wave-numbers
propagate at different speeds, so that localized initial data
would disperse (spread) in space.

In bounded systems, the set of normal wave modes becomes discrete.
For waves in a periodic $d$-dimensional cube with side $L$, the normal modes are
given by \eqref{NM} with a discrete set of wavenumbers
${\bm k} = \frac {2 \pi { \bm  l}}{ L}$ where ${ \bm  l} \in \mathbb{Z}^d$.
For different boundary conditions, normal modes of the linearized
problem may differ from the propagating plane waves~\eq{PW}.  For
instance, zero   boundary conditions in a rectangular box
 typically (but not always!) lead to  standing waves, 
\BE \label{SW}\psi(\B
r,t)=|A_\k|\sin ( {\bm k}\cdot{ \bm  r}+\varphi\sb{sp}) \, \sin (\o t+
\varphi\sb{t})\,,
 \ee
  where $\varphi\sb{sp}$ and $\varphi\sb{t}$ are
the space and time phases correspondingly. A more complex form of the
normal mode is given by   ocean planetary motions in a rectangular
domain $[0,L_x]\times[0,L_y]$ with zero boundary conditions, see
e.g. ~\cite{ped}: 
\be
\label{rectangular_mode}\psi(\B r,t)= |A_\k| \sin\!\Big(\pi\frac{mx}{L_x}\Big)
\sin\!\Big(\pi\frac{ny}{L_y}\Big)\, \sin\! \Big(\frac{\b}{2\o}x+\omega
t+\varphi\sb t \Big ) \,, \ee
\ese
where  $m$, $n \in\mathbb{N}$ are integers  and $
\omega=\beta/\big[2\pi\sqrt{(m/L_x)^2+(n/L_y)^2}\, \big]$ with a constant $\beta$
 called Rossby number.

\subsection{\label{sss:exp}  Equation of motion}
A rather general class of non-dissipative nonlinear waves can be described within the framework of the classical
Hamiltonian approach. This means that after a proper change of variables the motion equation in natural variables
(fluid velocity, electrical field, density variations, etc.) can be presented in the universal form of  canonical
Hamiltonian equations
 for canonical variables $b(\B r, t)$,    $b^*(\B r, t)$, which characterize the wave amplitudes. Here  ``~$^*$~"
 denotes complex conjugation.  The Hamiltonian equations for the space-homogeneous systems
 are most conveniently written
 in Fourier space, because it is a natural space for describing the wave solutions.
 Introducing the Fourier transform of  $b(\B r, t)$ and calling it $a_\k\= a (\k, t)$,
 the Hamiltonian equation  can
 be written   as follows~\cite{ZLF},
\be \label{HEM}
i\frac{d a_\k}{d t} = \frac{\partial {\cal H}}{\partial a_\k^*}\ .
\ee
Hamiltonian  $\C H\= \C H\{a_\k, a_\k^*\}$
is
usually (but not necessarily) is the energy of the wave system, expressed in the terms of the canonical
variables $a_\k$, $a_\k^*$ for all allowed by the boundary conditions wave vectors $\k$. In the simplest case
of a periodical box
${ \bm  k} = 2 \pi { \bm  l} / L$, with wavenumber ${ \bm  l} \in \mathbb{Z}^d$ and $L$  being
the box size  and $d$ is space dimension.

For the waves of small amplitudes (for example, when the elevation of the gravity waves on the water surface is
smaller then the wavelength) the Hamiltonian can be expanded   in powers $a_\k$ and $a_\k^*$:
\BSE\label{Ham}\bea {\cal H} &=& {\cal H}_2 + {\cal H}\sb{int}\,, \\
 {\cal H}\sb{int}&=&{\cal H}_3+ {\cal H}_4 +  {\cal H}_5 + \dots\,,
\label{Hexp}
\eea\ese
where $\C H_j$ is a term proportional to product of $j$ amplitudes $a_\k$ and the interaction
Hamiltonian $ {\cal H}\sb{int}$  describes the wave coupling,  as explained below.
We omitted here the independent of $a_\k$ and $a_\k^*$ part of the Hamiltonian
 $\C H_0$, because it does not contribute to the motion equation~\eq{HEM}. In this paper we consider only waves
 exited about steady equilibrium states,
i.e. if absent initially, the waves must remain absent for all time, $a_\k=a_\k^* \equiv 0$.
 Thus, the  linear Hamiltonian is zero, $\C H_1=0$.

Expansion~\eq{Hexp} utilizes the smallness of the wave amplitudes, therefore, generally speaking,
\BSE\label{small}
\BE \label{smallA}  {\cal H}_3>  {\cal H}_4 >  {\cal H}_5 > \dots\ .
\ee
In particular cases, due to specific symmetries of a problem, the odd expansion terms vanish
(i.g. for spin waves in magnetics with exchange interactions, Kelvin waves on quantum vortex lines).
In these cases, instead of~\eq{smallA} one has:
\BEA \label{smallB}  && {\cal H}_3 =  {\cal H}_5 =  {\cal H}_7=\dots =0\,, \\
 &&{\cal H}_4   >    {\cal H}_6 > {\cal H}_8 > \dots\ .
\eea\ese
Three-wave interactions often dominate in wave systems with small nonlinearity, e.g.
for Rossby waves in the Atmosphere and Ocean, capillary waves on the water surface, drift waves in plasmas,
 etc. On the other hand, if  ${\cal H}_3=0$, or if three-wave resonances are forbidden  (in the sense that will be clarified below) the leading nonlinear processes
  may be four-wave interactions.
  Further, there are examples of systems where the four-wave interaction
  is absent and the leading nonlinear process is five-wave, e.g. for one-dimensional
  gravity water waves \cite{zakh,5waves,5waves_K}, or even six-order, e.g. for  Kelvin waves on quantum vortex lines \cite{kozik_svistunov,6waves}. However, such higher-order wave systems are rather rare and,
  therefore, in this paper we will discuss  three- and four-wave interactions only, which describe most
    of weakly interacting waves.


\subsection{\label{sss:linear}  Non-interacting waves}
The first physically meaningful expansion term, quadratic   Hamiltonian
\BSE\label{Ham2}\be {\cal H}_2 = \sum_{n=1}^\infty \omega_\k |a_\k|^2\,,
\label{H2}
\ee
according to \Eq{HEM} produces a linear equation of motion,
 \BE \label{lin}
  i\, \frac{d a_\k}{d t}=\o_\k  a_\k\,,
 \ee\ese
 and thus describes noninteracting waves with the dispersion relation $\omega_\k \equiv \omega({  \bm  k})$.  For waves,
considered in this paper, when $\min_{a_\k, a^*_\k}\{\C H\}=\C H_0$,   $\omega_\k\ge 0$.  Notice, that $\C H_2$ in \Eq{H2}
does not have $a_\k a_{-\k}$ and $a_\k^* a_{-\k}^*$ terms. They were removed by   linear canonical transformation [known as
the  Bogolubov
 $(u,v)$-transformation] after which
  $\C H_2$ takes the fully diagonal form~\eq{H2}.
\subsection{\label{sss:3w} Three-wave   interactions}

First contribution to the interaction  Hamiltonian  $\C H\sb{int}$ is
 \BSE \label{H3}\be \label{H3a} {\cal H}_3=\frac12
\sum_{{ \bm  k_1}, { \bm  k_2},{ \bm  k_3}}  V^1_{23} a_{1}^* a_2 a_3\delta^1_{23}+\mbox{c.c.}\,,
 \ee 
 describes the processes of decaying of single wave
into two waves ($1\Rightarrow 2$ processes) or confluence of two waves into a single one  ($2\Rightarrow 1$ processes).
In \Eq{H3}
 for brevity we introduced notations $ a_1 \equiv a_{\k_1}$ etc. and $ \delta^1_{23} $
 is the Kronecker symbol, i.e
$\delta^1_{23} =1$ if and only if $ \k_1 + \k_2= \k_3.$
Clearly, $ V^1_{23}= V^1_{32}$. Generally speaking,  $\C H_3$ also includes  $a_1 a_2 a_3$ and $a_1^* a_2^* a_3^*$ terms that
describe
   $3 \Leftrightarrow 0$  processes (confluence of three waves or spontaneous appearance of three waves out of  vacuum).
   However they
 can be eliminated by corresponding nonlinear   transformation~\cite{ZLF} that leads to the canonical form of ${\cal H}_3$,
 presented in \Eq{H3}.

Hamiltonian $\C H_2+ \C H_3$ with \Eq{HEM} yields the three-wave equation:
\be\label{EM3W}
i\, \frac{d a_\k}{d t} = \o_\k a_\k + \sum_{{ \bm  k_1}, { \bm  k_2}} \Big[ \frac 12 V^\k_{12} a_{1} a_2 \delta^\k_{12}+
 V^{1 \, *}_{\k 2}  a_1 a_{2}^*  \delta^1_{\k 2}\Big ] \ .
\ee
\ese
Two sets of terms in the RHS of this equation  have time dependence of the form $\exp [-i (\o_2+\o_3)t]$ and
$\exp [-i( \o_2-\o_3)t]$ correspondingly [we used shorthand notations, $\o_j\= \o(\k_j)$].  They become important
if their frequencies are close to the eigenfrequency of $a_\k$, $\o_\k$:  $\o_2+\o_3\approx \o_\k$ or  $\o_2-\o_3\approx \o_\k$.
By relabeling  the wavevectors, we can write
both of these conditions in the same form as follows,
\BSE  \label{res3} \bea \label{res3a}
\o(\k_1)+\o(\k_2)&=&\o(\k_3)\ .
\eea
This condition of  time synchronization should be complemented by the condition of space synchronization that
formally originates from the Kronecker symbols in \Eq{EM3W},
\bea   \label{res3b}\quad \k_1 + \k_2&=& \k_3 \ .
\eea \ese
Both relations~\eq{res3} are named the resonance conditions of the three-wave interactions, or
conditions of the \emph{three-wave resonances}.

There exists a simple conditions for the three-wave resonance
conditions to be satisfied for the power-law dispersion relations
$\o \sim k^\alpha$ ($\alpha$=const). In 2D, it is most easily proved
graphically, as suggested in
\cite{vedenov}.
Thus, it was shown that the three-wave resonance is possible if and only if $\alpha \ge 1$
for the continuous case, $\k \in \mathbb{R}^2$.
Obviously, this
condition becomes a necessary condition if $\k$ is restricted to
discrete values due to boundary conditions.


\subsection{\label{sss:4w} Four-wave   interactions}
 When the three-wave resonances are forbidden, one has to account for processes with weaker nonlinearity, the four-wave interactions.   The canonical  part of the four-wave interaction Hamiltonian,
\bse\label{4}\be  {\cal H}_4= {1 \over 4}
\sum_{{ \bm  k_1}, { \bm  k_2},{ \bm  k_3}, { \bm  k_4}}
T^{12}_{34} a_{1}^* a_2^* a_3 a_4 \, \delta^{12}_{34} \,,
 \label{H4}
\ee
 describes a 4-wave scattering processes $2 \Leftrightarrow 2 $.
 Terms   $a_1a_2a_3a_4$
 and its complex conjugate describing $4\Leftrightarrow 0 $ processes
  can be   eliminated
 by an appropriate nonlinear canonical transformation \cite{ZLF}.  After that the four-wave interaction
 Hamiltonian takes the canonical
 form~\eq{H4}.
There also exist $1\Leftrightarrow 3$ systems with $a_1a_2a_3a_4^*$ and its complex conjugate
terms in  $\C H_4$ \cite{llnr09,ln_spectrum}. They can be treated similarly, but for simplicity we
omit them in the present paper.

  Note that besides trivial symmetries with  respect to the indexes permutations,
  $1 \leftrightarrow 2$
 and $3 \leftrightarrow 4$,  the interaction coefficient \eqref{4} has the symmetry
  $T^{34}_{12} = \big(T^{12}_{34}\big )^*$,
 because the Hamiltonian has to be real,  $\C H_4=\C H_4^*$.


The dynamical equation for the four-wave case follows from~\eq{HEM}
with the Hamiltonian $\C H=\C H_2+\C H_4$: 
\be\label{EM4W} i\frac{d
a_\k }{d t}= \o_\k a_\k +  \frac 12 \sum_{{ \bm  k_1}, { \bm  k_2},
{ \bm  k_3}} T^{\k1}_{23} a_{1}^* a_2 a_3 \delta^{\k1}_{23}\ . 
\ee\ese
Considering this equation similarly to~\eq{EM3W}, one realizes that
the terms in the RHS of~\Eq{EM4W} oscillate with the frequencies
$\o_2+\o_3-\o_1$ and becomes resonant if this combination is close to $\o_\k$.
in the other words, the condition of time synchronization (after proper
renaming of the variables) takes the form~\eq{res4a}
 \BSE  \label{res4}
\bea \label{res4a} \o(\k_1)+\o(\k_2)&=&\o(\k_3)+\o(\k_4)\,, \\
\label{res4b} \k_1 + \k_2&=& \k_3+\k_4\,, \eea \ese while \Eq{res4b}
represents condition of space synchronization that comes from the
Kronecker symbol in \Eq{EM4W}.



\subsection{\label{ss:Ex} Physical examples}

In the context of the problem at the hand, a choice of physically
important and methodologically illustrative Hamiltonian systems is
not an easy task. The corresponding wave systems should preferably
be well-studied, both theoretically and experimentally (or
numerically). They should be \emph{simple enough} to be understood
by the non-experts in the area of wave turbulence and at the same
time \emph{not too simple} in order to demonstrate the main
characteristics of the resonant wave systems described by different
nonlinear dispersive PDEs, with different number of interacting
modes and different boundary conditions.

\subsubsection{\label{sss:gw} Surface water waves}
Our first example is the system of surface water waves, with
 dispersion relation of the general form:%
  \bse\label{o-k}%
\be\label{o-kA} \o_k=\sqrt{g k+ \frac{\sigma\, k^3}{\rho}}\,, \ee%
where $g$ is the gravity acceleration, $\sigma$ is the surface
tension and $\rho$ is the fluid density.
 For small $k$ \Eq{o-kA} turns into  dispersion law for the gravity waves:
\be\label{o-kB}
\o_k=\sqrt{g k }\,,
\ee
while for large $k$ it is simplified to the capillary wave form
 \be\label{o-kC}
\o_k=\sqrt{\frac{\sigma\, k^3}{\rho}}\ . \ee 

In both limiting cases the dispersion law have scale-invariant form,
$\o_k\propto k^{\a}$. Notice that for the gravity waves $\a=\frac
12< 1$ and therefore the leading nonlinear processes are four-wave
scattering $2 \Leftrightarrow 2$ with the quartets as the primary
clusters, while for the capillary waves $\a=\frac 32$ and thus the
leading nonlinear processes are three-wave interactions of $2
\Leftrightarrow 1$ type. In this case the primary clusters are
triads. \ese

Surface water waves with the general dispersion law~\eq{o-kA} can be
described by the Hamiltonian equation of motion in the canonical
form~\eq{HEM}, that turns into \Eq{EM3W} for the capillary waves and
into \Eq{EM4W} for the gravity waves. Three-wave interaction
coefficient for the capillary waves reads as
\be V^1_{23} = {i \sqrt{\omega_{1} \omega_2
\omega_3}\over 8 \pi \sqrt{2 \sigma}}
\left[ {\C K_{k_2, k_3} \over  k_1\sqrt{ k_2 k_3 } } -
 {\C K_{k_1, -k_2} \over  k_3 \sqrt{ k_1 k_2 } }
-
 {\C K_{k_1, -k_3} \over k_2 \sqrt{ k_1 k_3 }  }
\right]\,,  \nonumber
 \ee 
 where
  \be
 \C K_{k_2, k_3}  = ({ \bm  k}_2  \cdot { \bm  k}_3) +  k_2 k_3\ .
\ee
The 4-wave interaction coefficient  for the gravity waves is given by rather long
expressions which can be found in \cite{zak_finite}.\\

\subsubsection{\label{sss:nls} Nonlinear Shr$\ddot{\hbox{o}}$dinger (NLS) model}

Probably the simplest known example of the four-wave systems are
waves in the nonlinear Shr$\ddot{\hbox{o}}$dinger (NLS) model of nonlinear optical
systems and Bose-Einstein condensates \cite{ZMR85,DNPZ}. {\em NLS
waves} have dispersion function and interaction coefficient as
follows:
\be%
\omega_k =k^2, \hspace{1cm} T^{12}_{34} = 1.\ee

\subsubsection{\label{sss:rossby} Rossby and drift waves}

Another important example of wave system
with dominating three-wave interaction, is {\em Rossby waves}, which are
similar to \emph{drift waves} in inhomogeneous plasmas. Their amplitudes can be
described
by the so-called  barotropic vorticity equation  which can be presented in the form similar to the canonical
 three-wave equation~\eq{EM3W}, but all $\k$'s taking values only in half of the Fourier space,
\bea
i\, \frac{d a_\k}{d t} = \o_\k a_\k + \hspace{4.5cm}
\nonumber \\
\!\!\!\! \sum_{{ k_{1x}},  k_{2x} \ge 0}
\Big[ \frac 12 V^\k_{12} a_{1} a_2 \delta^\k_{12}+
 V^{1 \, *}_{\k 2}  a_1 a_{2}^*  \delta^1_{\k 2}\Big ]; \;\;  (k_x>0) .
 \label{ReqA}
\eea
The phase space in this case is half of the Fourier space is a result
is because the original equation in the $x$-space is for a real variable (barotropic vorticity).
The difference in the Hamiltonian structure of the Rossby and
capillary waves yields  the difference in the form of the
conservation laws and therefore in their dynamical behavior. We will
discuss this later in greater detail.

Rossby waves on an infinite (or double-periodic)  $\beta$-plane have
dispersion function \cite{zakpit,balk_nazarenko}%
 \bse\label{Rossby-f}
 \be \label{Rossby-fA}\omega_\k
= {\beta \rho^2 k_{x}
\over 1 + \rho^2 k^2}  \,, 
\ee %
 where $\rho = \sqrt{gH}/f$  is the Rossby deformation radius
$H$ is the fluid layer thickness,
$f= 2 \Omega \sin \theta$ is the Coriolis parameter, $\theta$ is
the latitude angle ($\beta$-plane approximates a local region
 on surface of a rotating planet), $\Omega$ is the planet rotation
 frequency and $\beta$ is the gradient of the Coriolis parameter,
 $\beta = 2 \Omega \cos \theta /R$, and $R$ is the radius of the planet.

In the case of zero boundary conditions in a plane rectangular
domain (\emph{oceanic Rossby waves}), the form of the eigenmode is
given by Eq.(\ref{rectangular_mode}), corresponding dispersion
function has the form%
\be \label{Rossby-fB}\o_k = {\beta L \over 2 \pi \sqrt{m^2+n^2}} .   \ee %
%
Note that this dispersion relation
coincides with relation  \Eq{Rossby-fA} in the limit $\rho \to \infty$
taking into account that $ (k_x,k_y) = (\beta /2 \o \pm \pi m/L, \, \pm \pi n/L)$
[which follows from Eq.(\ref{rectangular_mode})].
However, the resonant mode sets are different because the resonance in ${\bm k}$
is now replaced by the resonance conditions in $m$ an $n$.

One more example is  \emph{Atmospheric Rossby waves}, propagated  on a rotating
(with angular velocity $\Omega$)
sphere. Eigenmodes in this case,
$ \displaystyle  Y_\ell^m (\sin \varphi,\lambda ) \exp\Big[\frac{2i m}{\ell(\ell+1)}
t\Big]$, are proportional to the spherical functions $Y_\ell^m $,
 where   $\ell \le 1$ and $|m|\le \ell $ are integers  and    $\
\varphi  $ and $\ \la \ $ are latitude and longitude
correspondingly.
 In this
case dispersion function is of the form%
\be \label{RossbyS} \o_{\ell,m} =
{2\, m \Omega  \over [\ell(\ell+1)]}\ .
\ee
\ese%

Notice that difference in the dispersion relations~\eq{Rossby-f} leads to essential difference
in the topology of resonant clusters, and consequently to essential difference
in the dynamical  and statistical behavior of the systems.

For concreteness we present here the interaction coefficients of the Rossby waves in the (infinite or
double-periodic) $\beta$-plane \cite{98Pit}:%
\be \label{RossbyInfCoeff}   %
V^1_{23} = { \beta \sqrt{| k_{1x} k_{2x} k_{3x} | } \over 4 \pi i}
  \left[  { k_{1y} \over 1 + \rho^2 k_1^2} - { k_{2y} \over 1 +
\rho^2  k_2^2}  - { k_{3y} \over 1 + \rho^2 k_3^2} \right] \! . \nonumber
 \ee

The interaction coefficients for the atmospheric Rossby waves can be
found in \cite{sil,ped,KPR} and for oceanic Rossby waves - in
\cite{KR}.

\section{\label{ss:DvsS} Regimes of finite-size Wave Turbulence}

What happens when, due to the finite size, the number of exact resonances and
active quasi-resonances
is depleted or absent?
The finite-size effects in WT can be characterized by considering the
 nonlinear frequency broadening $\Gamma$ (i.e. the inverse  the characteristic time
 of nonlinear evolution) and comparing it to  the frequency spacing $\Delta_\omega$
 between the finite-box
 eigenmodes. For simplicity, we will restrict our attention to the periodic boundary
conditions, in which case
\BE
\label{deltaW}
\Delta_\omega = \left|\frac{\partial \omega_k}{\partial {\bm k}} \right|\frac{2\pi}{L}
\sim \frac{ \omega_k}{k L}.
\EE
``Twiddle'' here means that this is an order of magnitude relationship, which corresponds the
approximate character of the physical estimates given below.


The kinetic equation  is applicable when $\Gamma \gg \Delta_\omega$,
 this is the \em kinetic regime. \em
A qualitative different behavior can be expected in  the opposite limit $\Gamma \ll \Delta_\omega$: this is a
regime of
 \em discrete wave turbulence. \em
These two regimes are realized when WT forcing is rather high (but not too high so that the nonlinearity
is still weak) and low respectively.
However, we will also see that there is also a rather wide intermediate range of forcing for
which there is a regime with $\Gamma \sim \Delta_\omega$, which we will call \em mesoscopic wave turbulence. \em

 Name \em mesoscopic \em refers to an observation made  in  \cite{z2} and \cite{Naz-2006}
 that in existing numerical simulations of the gravity water waves there may be
regimes where the statistical properties of the infinite-box systems coexist with effects due to
the $\bm k$-space discreteness associated with a finite computational box. In was further argued in
\cite{slavedMHD} (in the context of MHD turbulence)
that such a mesoscopic regime is active in a \em wide  intermediate range of wave amplitudes. \em
The key reason for such a wide mesoscopic range is the fact that  the typical values of $\Gamma$ for
the discrete (dynamical) and the kinetic (statistical) regimes are typically strongly separated.

\subsection{\label{sss:onset} Discrete turbulence (small box, weak waves)}

 In the discrete WT regime, when $\Gamma \ll \Delta_\omega$, only the terms in the dynamical
 equations which corresponding
 to  \em exact \em wavenumber and frequency resonances contribute to the nonlinear wave dynamics.
 All the other terms rapidly oscillate and their net long-term effect is null.
 The most clear example here is the case when there is no exact resonances, like in the system
  of the capillary water surface waves.
 In this case, the averaged (over the fast linear oscillations) nonlinearity is negligible
 and the turbulent cascade over scales is arrested.
 One can see an analogy with KAM theory which says that trajectories of a
 perturbed (in our case by nonlinearity) Hamiltonian system remain close to
 the trajectories of the un-perturbed integrable system (in our case the linear wave  system
 whose trajectories are just harmonic oscillations of the individual modes)
 if there is no resonances. Of course, this analogy should be taken with caution because
 even in absence of the lower-order resonances (e.g. triad resonances for the capillary waves)
  higher-order resonances may be important.

  Thus, for the discrete WT regime we have the following reduced dynamical equations
 \BSE
 \BE
 \label{disc3wave}
 i \frac{d a_k}{dt} = \sum\limits_{1,2} \left( \frac 1{2} V_{12}^{k} a_1 a_2  R_{12}^{ k}
 + V_{1k}^{2*} a^*_1  a_2 R_{1k}^2 \right),
 \EE
 for the three-wave case [~\Eq{EM3W}
 in which we retain only exact wave resonances]
 and
\begin{equation}
i \frac{d a_k}{dt} =  \frac 1{2} \sum\limits_{
{1,2,3}} {W}_{3k}^{12} a_1 a_2 a_3^* \,  R_{3k}^{12}
\ ,
\label{disc4wave}
\end{equation}
\ese
for the four-wave case [ \Eq{EM4W} with only exact  resonances left].

In equation \eqref{disc3wave}, factor $R_{12}^{ 3}$  is equal to one
when modes ${\bm k}_1, {\bm k}_2$ and $ {\bm k}_3$
are in exact wavenumber and frequency resonance, and it is zero otherwise.
Respectively in \eqref{disc4wave},
$R_{34}^{12}$ is equal to one  when modes ${\bm k}_1, {\bm k}_2, {\bm k}_3$ and $ {\bm k}_4$
are in exact wavenumber and frequency resonance, and it is zero otherwise.

Some resonant triads/quartets (if at all present) may be isolated, in which
case their dynamics
is integrable, and the respective nonlinear oscillations can be expressed in terms
of the elliptic functions. Some  triads/quartets may be linked and form clusters
of various sizes, whose dynamics is more complicated and to some extent may be chaotic,
especially for larger clusters.
Study and classification of such exact resonances and their
clusters was initiated in
\cite{KPR,kar90,kar91,KR}
developed further in many papers including
\cite{Naz-2006,NazarLvov2006,KM07,KL-07,KL-08,LPPR,KNR-08}.
Examples of small and large clusters for the Rossby waves (three-wave system) can be found in
\cite{KL-07,KL-08,LPPR} and for the gravity water waves (four-wave system) in
\cite{KNR-08}.
%

Frequency broadening  $\Gamma$ for the discrete WT  can be estimated from
the dynamical equations \eqref{disc3wave} and \eqref{disc4wave},
\BSE\label{DO}\BEA
\label{DOa}
\Gamma = \Gamma^{(3w)}_{D}&\simeq & | V
 a_k|\,{\mathcal N} ,
 \\ \label{DOb}
\Gamma = \Gamma^{(4w)}_{D}&\simeq & | W
  a_k^2|\,{\mathcal N} ,
\EEA\ese
where $V= V^{k}_{12}$ and $W=W ^{12}_{3k}$ are  the interaction coefficients
in \eqref{disc3wave} and \eqref{disc4wave} respectively.
Subscript $ {D}$ indicates that this is a discrete-regime estimate, and
superscripts $3W$ and $4W$ stand for "three-wave" and "four-wave" respectively.
 Here ${\mathcal N} $ is the number of exact resonances which are dynamically
 important at a fixed ${\bm k}$, which less or equal to
  the number of modes connected to ${\bm k}$ in the resonant
  cluster. For simplicity we assumed that that all the dynamically important
   resonances are local, i.e. $k_1 \sim  k_2 \sim k_3  \sim k $.
Strictly speaking,
  the estimates \eqref{DOa} and  \eqref{DOb} are only valid if
  ${\mathcal N} $
  is not too large,
  because when ${\mathcal N} \gg 1$ one should expect statistical cancelations
  of the effect of different triads/quartets, and our estimates would have to be
  modified. This is the case for the example
  of MHD turbulence considered in \cite{slavedMHD}.
 Also, our estimates would have to be modified
 for systems with nonlocal in ${\bm k}$ interactions.

Thus, the condition of the discrete turbulence regime,  $\Gamma_{D}\ll \Delta_\omega $,
becomes
\BSE\BEA
\label{frLow1}
| V
 a_k| & \ll & \frac{ \omega_k}{k L {\mathcal N}}, \quad \mbox {for 3-wave systems,}
 \\
 \label{frLow2}
 | W
  a_k^2| &\ll & \frac{ \omega_k}{k L {\mathcal N}}, \quad \mbox {for 4-wave systems}.
\EEA\ese

\subsection{\label{sss:KE} Kinetic wave  turbulence (infinite-box limit)}

The kinetic regime comprises the classical infinite-box weak WT theory,
which is reviewed in the Appendix to this paper, including recent
theory extensions to description of the higher-order wave moments
and PDF and finding solutions
corresponding to turbulence intermittency \cite{cln1,cln2,cln3}.
In this regime, the frequency resonance broadening, denoted $\Gamma_{K}$,
is
determined by the kinetic equation \eqref{ke3} for
for three-wave systems and \eqref{ke4} for the four-wave systems.
This gives for
 $\Gamma_{K}$:
\BSE \BEA
\label{Ka} 
\Gamma^{(3w)}_{K}  \simeq  |V|^2 n_k k^d \big/ \omega_k
  \simeq  | V|^2 |a_k|^2 (kL)^d \big/ \omega_k
   \,, ~~~~ \\ 
\Gamma^{(4w)}_{K} \simeq |W|^2 n_k^2 k^{2d} \big/  \omega_k
  \simeq   | W|^2 |a_k|^4 (kL)^{2d} \big/ \omega_k    \ ,  \label{Kb}
\EEA\ese
where, for simplicity, we have assumed that the wave spectrum is not too narrow
and the range of wavenumbers interacting with ${\bm k}$ is of width $\sim k$.

The upper bound for applicability of the wave kinetic equations follows from the
condition of weak nonlinearity $\Gamma_{K}\ll \omega_k $ which gives
\BSE \BEA \label{Ka11} 
  | V
  a_k | (kL)^{d/2} &\ll &\omega_k   \,,   \quad \mbox {(3-wave),} \\ \label{Kb11}  
  | W|
 |a_k|^2   (kL)^{d} & \ll &  \omega_k     \,,   \quad \mbox {(4-wave).}
\EEA\ese
Also,  the wave amplitudes should be large enough
for the  broadening $\Gamma_k$ to be much greater
than the frequency spacing $\Delta_\omega$. Together with
 (\ref{deltaW}), this condition gives
\BSE \BEA \label{1Ka} 
  | V  a_k |    \gg    \frac{\omega_k}{(kL)^{(d+1)/2}} ,   \quad \mbox {for 3-wave systems,} \\ \label{1Kb}  
  | W||a_k|^2      \gg      \frac{\omega_k}{(kL)^{d+1/2}} ,   \quad \mbox {for 4-wave systems.}
\EEA\ese %


\subsection{\label{sss:meso} Mesoscopic turbulence and sandpile dynamics}

Consider first the case when the number of connections  of mode
$\bm k$ in its discrete resonant cluster is relatively small,
 ${\cal N} \gtrsim 1$, as it is the case, e.g.,
for  the case of the gravity water waves.
Comparing the range of kinetic WT, \eqref{1Ka}, \eqref{1Kb}, and the one of discrete WT,
  \eqref{frLow1}, \eqref{frLow2}, one can see
that  there exists a gap,
\BSE \BEA \label{1Ma} 
\frac 1 {k L {\mathcal N}} \gg   \frac{ | V a_k | }{\omega_k}  \gg    \frac 1 {(kL)^{(d+1)/2}} ,   ~~\quad \mbox {(3-wave),} \\ \label{1Mb}  
\frac 1 {k L {\mathcal N}}  \gg   \frac{| W||a_k|^2 }{\omega_k} \gg       \frac 1 {(kL)^{d+1/2}} ,   \quad \mbox {(4-wave).}
\EEA\ese %
 in which both the conditions for the kinetic WT and for the discrete WT are satisfied.
This means that in the region~\eqref{1Ma},~\eqref{1Mb} the wave behavior
is neither pure discrete nor  pure kinetic
WT. Existence of such a gap was first pointed out in
\cite{slavedMHD} in the context of MHD wave turbulence.
Region~ \eqref{1Ma},~\eqref{1Mb} possess the features of both types of turbulent
behavior described above.  In the other words, in this region both types
of WT may exist and the system may oscillate in time (or parts of the ${\bm k}$-space) between the two
regimes  giving rise to a
qualitatively new type of WT: \emph{mesoscopic
wave turbulence}. It was suggested in \cite{Naz-2006} (in the
context of the surface gravity waves) that in forced wave systems the
discrete and the kinetic
 regimes may alternate in time, see figures~\ref{sand}.
  Namely, let us consider WT with initially very weak or zero intensity,
  so that initially WT is in the discrete regime, and let us permanently
  supply more wave energy via a weak source at small $k$'s.
 During the discrete phase (with fully or partially arrested cascade)
 the wave energy accumulates until
 when the resonance broadening $\Gamma_D$ becomes of order of the frequency spacing $\Delta_\omega$. After
 that the turbulence cascade is released to higher $k$'s  in the form of an
 "avalanche" characterized by predominantly kinetic interactions.
 At the moment of triggering the avalanche, the broadening $\Gamma$ jumps up
 from $\Gamma =\Gamma_D$ to $\Gamma = \Gamma_K \gg \Gamma_D$, see the upper figure~\ref{sand}.
 In the process of  the avalanche release, the mean wave amplitude lowers so that
 the value of broadening $\Gamma =\Gamma_K$ becomes of order of the frequency spacing $\Delta_\omega$.
 Remember, for not too large ${\mathcal N}$ in this intermediate range $\Gamma_K \gg \Gamma_D$.
 Thus, at this point
 the system returns to the energy accumulation stage in the discrete WT regime,
 and the cycle repeats, see figures~\ref{sand}. Because of the obvious analogy, this scenario was called
  \em sandpile behavior \em in \cite{Naz-2006}.

As we see, the sandpile behavior is characterized by a \em hysteresis \em
 where in the same range of amplitudes,
from $A_1$ to $A_2$ in the lower figure~\ref{sand},
the WT intensity increases in the discrete regime and decreases in the kinetic regime.
\begin{figure}
  \includegraphics[width=9cm]{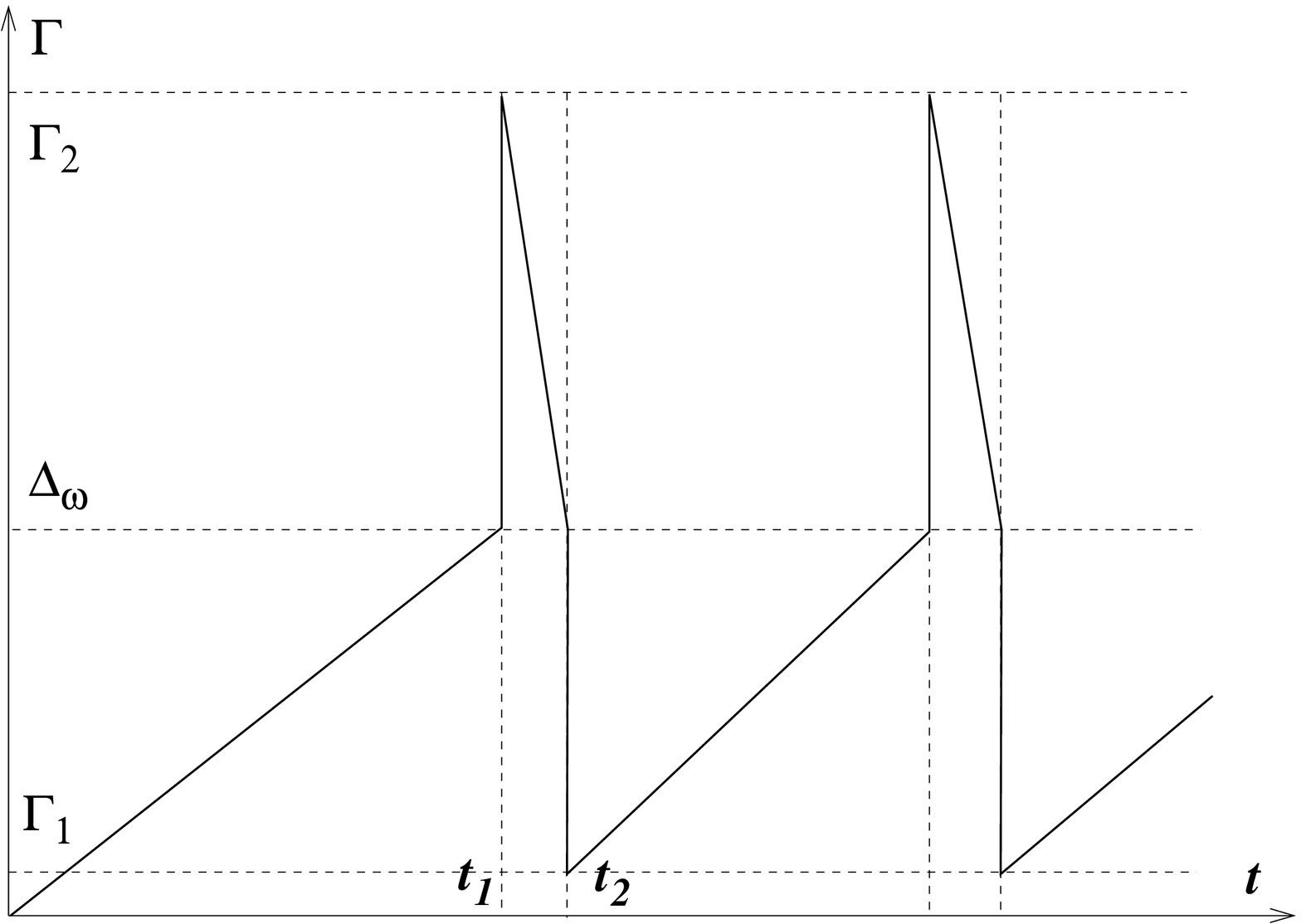}\\
   \includegraphics[width=9cm]{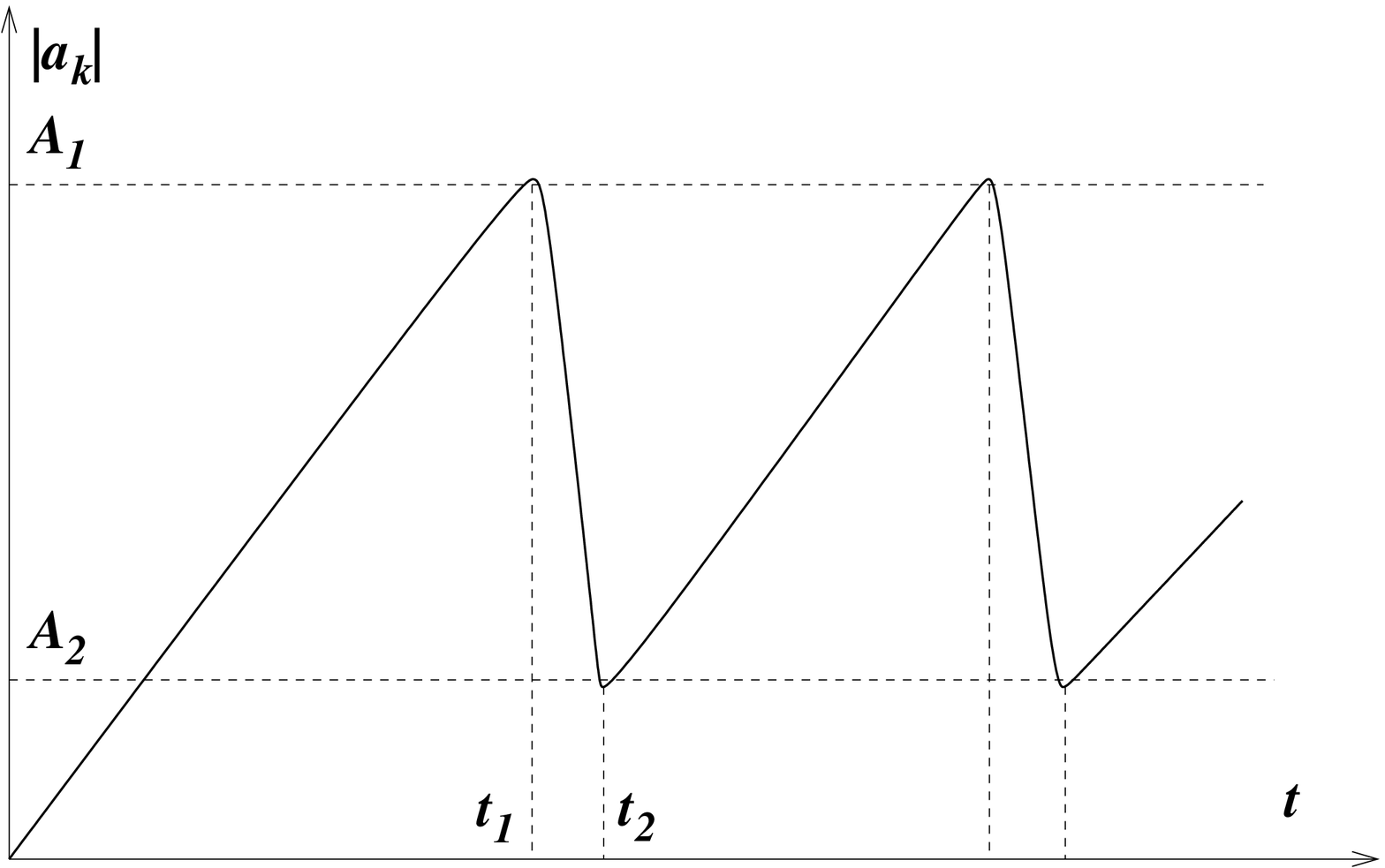}\\
  \caption{"Sandpile" behavior in wave turbulence. Upper graph: the frequency broadening
  $\Gamma$ follows the discrete turbulence dependence $\Gamma =\Gamma_D$
  until reaching the value $\Gamma =\Delta_\omega$ at time $t=t_1$, at which point it jumps
  to the kinetic branch $\Gamma =\Gamma_K \gg \Gamma_D$ and rapidly drops in the kinetic
  regime to the value $\Gamma =\Delta_\omega$ at time $t=t_2$. Then it jumps
  back to the discrete branch $\Gamma =\Gamma_D \ll \Gamma_K$, after which the cycle repeats.
  Lower graph:
  the amplitude gradually grows to $a_k \sim A_1$  for $t<t_1$ and
  then quickly drops to $ A_2$ for $t_1 <t<t_2$, after which the cycle repeats.
   For the three-wave systems
  $A_1 \sim (\omega /V) (kL)^{-(d+1)/2}$ and $A_2 \sim \omega/(kL V {\mathcal N})$
  and for the four-wave systems
   $A_1 \sim (\omega /W)^{1/2} (kL)^{-(2d+1)/4}$ and $A_2 \sim \sqrt{\omega/(kL W {\mathcal N})}$.
     } \label{sand}
\end{figure}

 For the small-amplitude part of the sandpile cycle, the
  system will be close to the critical spectrum,
  where resonance broadening $\Gamma_K$ is of order of the omega-spacing $\Delta_\omega$.
 This gives the frequency spectrum  $\omega^{-6}$, which was predicted in
 \cite{Naz-2006} and experimentally
 confirmed in \cite{denis}   (c.f.  $\omega^{-4}$ for the KZ
 spectrum in this case \cite{zakh_fil}).
 Finding spectrum close to the large-amplitude part of the cycle is not so
 straightforward because we do not know the dependence of ${\mathcal N}$ on
 $\omega$.

 So far, we only considered the case when ${\mathcal N}$ is not too large.
 Case ${\mathcal N} \gg 1$ can be very different.
 Namely, instead of the range where \em both \em conditions satisfied simultaneously,
 the one for the kinetic WT, \eqref{1Ka}, \eqref{1Kb}, and the one for the discrete WT,
  \eqref{frLow1}, \eqref{frLow2},
  one gets a range where \em none \em of these two conditions are satisfied.
  This kind of mesoscopic turbulence was considered using the MHD
  example in \cite{Naz-2006}.
  We will see that in this case the frequency broadening $\Gamma$ remains
  of the order of the omega-spacing $\Delta_\omega$ in a broad  (mesoscopic)
  range of wave amplitudes. Remembering that $\Gamma$ is a characteristic
  nonlinear evolution time, we note that constancy of $\Gamma$ points at
  a possibility that the energy transfer in such a mesoscopic regime is
  driven by a hidden effectively linear process, which is yet to be understood.

\subsection{\label{sss:reg} Possible coexistence of different regimes}

Strength of WT typically varies in along the turbulent cascade in the
${\bm k}$-space and, therefore, one may expect different wave turbulence regimes present in the
different parts of the ${\bm k}$-space at the same instant in time.
For example, the nonlinearity increases along the cascade toward high wavenumbers in
 WT of surface gravity waves and of
  MHD Alfv\'{e}n waves. Thus we can expect WT in these systems to
 be discrete at low $k$'s and kinetic at high $k$'s.
 Moreover, on the cross-over regions one can expect nontrivial gradual transition
which involves blending and interaction of different dynamical and statistical mechanisms.
 This effect is expected to be more pronounced if the interaction of scales is
 nonlocal, so that some wavenumber(s) from a particular resonant triad (or quartet) could be
 in the discrete range whereas the other wavenumber(s) from the same triad (or quartet)
 could be in the kinetic range.
 As a result, in the cross-over range a continuous spectrum described by the kinetic
 equation (e.g. KZ) could coexist with selected few  modes belonging to isolated resonant clusters
 which would evolve coherently at deterministic timescales.
Moreover, the same set of modes might randomly
alternate in time from being discrete to kinetic and back,
as we described above in the sandpile scenario.

Some basic consequences of variability of the finite-size effects
in the $\bm k$-space can be seen in an very simple
kinematic cascade model  suggested in \cite{conn}.
This model builds a ``cascade tree'' in the following three steps.
\begin{itemize}
\item
 Let us put some energy into a small collection of initial
modes. We denote this initial collection of excited modes by
$S_0$ (e.g. in within a circle or a ring at small $k$'s which
corresponds to forcing at large scales). One can view set $S_0$
as the cascade tree's ``trunk''.
\item
 Next, find the modes which can interact with the initial ones at the
given level of nonlinear broadening $\Gamma$. Namely, we define a new set
of modes
$S_1$ as the union of all $k$'s satisfying the quasi-resonance
conditions,
\BSE \BEA
| \omega_{3} - \omega_{2} - \omega_{1} | < \Gamma\,, \quad
{\bm k}_3 - {\bm k}_2 - {\bm k}_1 =0
\EEA
 for the 3-wave case and
 \BEA
| \omega_4 +\omega_3- \omega_2 - \omega_1 |& < &\Gamma, \nonumber \\
{\bm k}_4 +{\bm k}_3 - {\bm k}_1 - {\bm k}_2 & = & 0,
\EEA\ese
 for the 4-wave case,
 with all but one wavenumbers in $S_0$ and the remaining
wavenumber outside of $S_0$.
 Provided that $\Gamma$ is large enough,
the set $S_1 $ will be greater than $S_0$.
Set $S_1$
comprises the cascade tree's ``biggest branches''.
\item
 Now iterate this procedure to generate a series
of cascade generations $S_0$, $S_1, ...$, $S_N$ which will
mark  the sets of active modes as the system evolves.
The union of these sets constitutes the whole of
 the cascade tree with all of its bigger and smaller branches included.
\end{itemize}

This model is purely kinematic. It does not say anything
about how energy might be exchanged dynamically among
the active modes, or how rapidly a certain cascade generation is reached.
However,  the kinematics
alone allows one to make some interesting observations about the systems
with variable in $k$ finite-size effects.

Let us consider the example of the gravity waves on deep water, for which
the following results were obtained in \cite{Naz-2006}.
If one starts with a set of low-$k$ modes,
with broadening $\Gamma$ below a critical value $\Gamma\sb{crit} = 1.4 \times 10^{-5}$,
a finite number of modes outside the
initial region get excited  (generation
2) but there will be no quasi-resonances to carry energy to outer
regions in further generations. If the broadening is larger than
$\Gamma\sb{crit}$, the energy cascades infinitely.
Further, such the kinematic cascades were shown to have the fractal snowflake structure
with  the active
modes being rather sparse in the front of the cascade propagating
to higher $k$, with pronounced anisotropic and intermittent
character.

Similar picture of intermittent cascades was also observed for the
capillary wave system \cite{conn}. However, because there is no
exact resonances for this system, the generation 1 an higher appear
only if $\Gamma$ is greater than some minimal value $\Gamma\sb{crit1}$.
Further, there exists a second critical value $\Gamma\sb{crit2} > \Gamma\sb{crit1}$:
the number of generations is finite for $\Gamma\sb{crit2} > \Gamma > \Gamma\sb{crit1}$
and the cascade process dies out not reaching infinite $k$'s,
whereas for $\Gamma > \Gamma\sb{crit2}$ the number of generations is
infinite and the cascade propagates to arbitrarily high $k$'s.
Note that the later property makes the capillary wave system
different from the gravity waves for which the cascade always
spread through the wavenumber space infinitely
 provided $\Gamma > \Gamma\sb{crit}$.

 Another example where the (three-wave) quasi-resonances and the
  kinematic energy cascades were studied
 is the system of inertial waves in rotating 3D fluid volumes \cite{lidya}.
 This system is anisotropic and the study of the kinematic cascades
 allows to find differences between the 2D modes, with wavevectors perpendicular
 to the rotation axis, and the 3D modes.
It appears that the "catalytic" interactions which involve triads including simultaneously
2D and 3D wavevectors dominate over the triads which involve 3D wavevectors only.

\section*{Discussion}\label{s:discussion}

In this paper we have considered the three different regimes which can be observed in wave turbulence (WT) bounded by a finite box
 \em -- discrete, mesoscopic and kinetic. \em For very low amplitudes and small boxes, we expect the \em discrete WT, \em
  whose dynamics is driven by the exact resonances. In the opposite infinite-box limit, we expect the \em kinetic WT,
  \em
   which is driven  by quasi-resonances and for which the exact resonances do not play a role as they are hugely outnumbered by the quasi-resonances. This is the classical and the most studied WT regime, and it is summarized in our Appendix.
In the middle, there is a regime of the \em mesoscopic WT. \em We have shown that
this regime is characterized by
sandpile-like  oscillations between the discrete and the kinematic regimes (if the size of the active resonant clusters is small)
or it settles to an intermediate (critical) state in which the nonlinear frequency broadening is of order of the frequency spacing
between the discrete modes for a wide range of wavenumbers (if the size of the active resonant clusters is large).

The key fact that has led us to the observation that the mesoscopic regime should be realized in a \em wide \em
range of wave intensities, is that the dependence of the frequency broadening on the wave intensity is
very different for the dynamical and the kinetic equations; cf $\Gamma_D$ given by (\ref{DOa}), (\ref{DOb}) and
$\Gamma_K$ given by (\ref{Ka}), (\ref{Kb}). Thus,  for the same
 wave intensities in which $\Gamma_D$  and
$\Gamma_K$ are typically very different in size, and there exist a wide mesoscopic range where either \em both \em
the discrete and the kinetic regimes can exist, or \em none \em of them is realizable, - hence the two types
of the mesoscopic behavior described above.

Signs of bursty  behavior typical of the sandpile behavior suggested in this paper has already seen in
laboratory and numerical experiments \cite{denis,Naz-2006}. In future, one should aim to
perform more direct diagnostics of the quantities allowing to identify and to distinguish the
different WT regimes describe in the present paper,  including the nonlinear frequency broadening and
character of its evolution in time.

\appendix
 \section{\label{s:KWT}Kinetic  wave turbulence}

Classical Wave Turbulence theory provides a statistical description
of weakly nonlinear waves with random phases. As discussed above,
theory of wave turbulence  is valid in a range of wave-field
strengths such that %
\be 1 > {\Gamma  \over \o _k} >  {\Delta_\omega \over
\o _k} \sim \frac 1{kL}, \label{wt_range} \ee %
where $\Gamma $  is given by (\ref{Ka}) or (\ref{Kb}) for the
three- or  four-wave processes respectively.

The most popular statistical object in the theory of wave turbulence  is the waveaction spectrum,
although theory of wave turbulence
has been recently extended to description of higher moments and probability
density functions (PDF) in \cite{cln1,cln2,cln3}. This allowed to deal with non-gaussian
wave fields, as well as to study validity of the underlying statistical assumptions such as \emph{e.g.}
random phases.  We will now briefly describe these results.

Let us represent the complex amplitudes as $a_k = \sqrt{J_k} \psi_k $
with wave intensity $J_k \in \mathbb{R}^+$ (positive real number) and phase factor $\psi_k
\in \mathbb{S}^1$ (complex number of length 1).
Let us define the $M$-mode joint PDF ${\cal P}^{(M)}$ so that
the probability for the wave intensities of the selected $M$ modes, $J_k $, to be in the
range $(s_k, s_k +d s_k)$ and for their phase factors $\psi_k$ to be on
the unit-circle segment between $\xi_k$ and $\xi_k + d\xi_k$ is
${\cal P}^{(M)} \, \prod_{k=1}^M ds_k \, |d\xi_k|$.
(Therefore ${\cal P}^{(M)}$ is a function of $2M+1$ variables: $M$ amplitudes, $M$ phases and time).

Notion of random phases refers
to the cases where all factor $\psi_k$ are statistically independent and uniformly
distributed on $\mathbb{S}^1$, i.e.
\be
{\cal P}^{(M)} =  {1 \over \left( 2 \pi  \right)^M }\, {\cal P}^{(M)}_a
\ee
for any $M \le N$, where $N$ is the total number of dynamically active modes.
Here ${\cal P}^{(M)}_a$ is the joint PDF of the amplitudes only.
Kinetic WT considers wavefields with random phases at some initial
time and with intensities satisfying condition (\ref{wt_range}). This leads to the
following equation for the joint PDF for the three-wave case,
\bea
\frac{\partial {\cal P}^{(N)}}{\partial t} &=& { 16 \pi  }
\int |V^{1}_{23}|^2\delta(\o _1-\o _2 - \o _3 )
\delta({ \bm  k}_1 -{ \bm  k}_2-{ \bm  k}_3) \nonumber \\
&& \hskip - 1cm\times \left[{\delta  \over \delta s} \right]_3 \left(s_1 s_2s_3
\left[{\delta \over \delta s} \right]_3 {\cal P}^{(N)} \right)  \,
 d { \bm  k}_1 d { \bm  k}_2
 d{ \bm  k}_3\,,
\label{peierls3} \eea where $\displaystyle  \left[{\delta  \over \delta s}
\right]_3 = {\delta  \over \delta s_1}-{\delta  \over \delta s_2} -
{\delta  \over \delta s_3}. $ This equation was first derived for a
specific example of waves in anharmonic crystals by Peierls
\cite{peierls} and for general three-wave systems in
\cite{cln1,cln3,newell_jakobsen}. It was also extended to the four
wave systems in \cite{cln2}.
Note that the phase variables are not involved in these equations. Therefore, the random
phase assumption is consistent with these equations, namely the system which has random phases
initially will remain random-phased over the typical nonlinear time (i.e. its PDF will remain
independent of $\xi$'s). Thus, equations for the joint PDF (\ref{peierls3})
allows an {\em a posteriori} justification of the random phase assumption underlying their derivations.

However, as we already mentioned, the most frequently considered
object in the theory of wave turbulence  is the spectrum which is
defined as%
\be n_k = \left( {2 \pi  \over L } \right)^{d} \langle J_k
\rangle, \ee%
where $d$ is the dimension of the space and the angular brackets
mean the ensemble averaging over the wave statistics. The spectrum
is a one-mode statistical object, and it is the first in
the series of one-mode moments, %
\be M_k^{(p)} = \left( {2 \pi  \over L } \right)^{pd} \langle J_k^p
\rangle
= \left( {2 \pi  \over L } \right)^{pd}
\int_0^\infty s_k^p {\cal P}^{(1)} (s_k) \, ds_k. \nonumber \ee %

Note that for deriving closures for the one-mode objects the random phase property is insufficient
and one has to assume additionally that the amplitudes $J_k$ are also statistically independent
of each other at different $k$'s.
Statistical independent of the amplitude can also be justified based on the
equation for the joint PDF (\ref{peierls3}),
although this issue is
more subtle than the phase randomness because variables $s_k$ do not separate
in (\ref{peierls3})
and, therefore any product factorization of the
joint PDF in terms of the one-mode PDF's would not generally be preserved by the nonlinear
evolution. However, this situation seems to be typical for many systems, \emph{e.g.} for the
relation between the multi-particle and one-particle distribution functions described by the
Louisville and Boltzmann equations respectively. In these situations, a sufficient for the closures
property is that the {\em low-order} PDF's, ${\cal P}^{(M)}$ with $M \ll N$,
 can be product factorized. It can be seen from (\ref{peierls3})
 that it is the case for the weakly nonlinear wave systems, i.e. that factorization,
 $
 {\cal P}^{(M)} = \prod_{k=1}^M {\cal P}^{(1)}_k + O(M/N),
$
survives over the characteristic nonlinear time.

Importantly, the distribution of wavefields in the theory of kinetic WT  does not
need to be Gaussian or close to Gaussian, and one can consider evolution of the one-mode
PDF's ${\cal P}^{(1)}$ that correspond to strongly non-gaussian fields (Gaussian
fields would mean ${\cal P}^{(1)} \sim e^{-s/\langle J \rangle} $).
 Integrating the joint PDF equation (\ref{peierls3})
we get
\begin{equation}
{\partial  {\cal P}_k^{(1)}\over \partial t}+  {\partial F_k \over \partial s_k}  =0,
 \label{pa}
\end{equation}
 with $F$ is a probability flux in the s-space,
\begin{equation}
F_k=-s_k \Big(\gamma_k {\cal P}_k^{(1)} +\eta_k {\delta {\cal P}_k^{(1)} \over \delta s_k}\Big)\ .
\label{flux1}
\end{equation}
where for the three-wave case we have:
\begin{widetext}\bea
 \eta_k & =& 4 \pi  \int
\big(|V^k_{12}|^2
\delta(\o _k -\o _1-\o _2 )
\delta({ \bm  k} -  { \bm  k}_1 -{ \bm  k}_2)
+2 |V^2_{k1}|^2
\delta(\o _2 -\o _k-\o _1 )
\delta({ \bm  k}_2 -  { \bm  k} -{ \bm  k}_1) \big)
  n_{1} n_{2}
\,  d { \bm  k}_1 d { \bm  k}_2  ,  \label{RHO} \\ \nonumber
 \gamma_k & = &
8 \pi  \int
\big(
|V^k_{12}|^2
\delta(\o _k -\o _1-\o _2 )
\delta({ \bm  k} -  { \bm  k}_1 -{ \bm  k}_2) \, n_1 +
 |V^2_{k1} |^2 \Delta^2_{k1}
 \delta(\o _2 -\o _k-\o _1 )
\delta({ \bm  k}_2 -  { \bm  k} -{ \bm  k}_1) (n_1-n_2)\big)
\,  d { \bm  k}_1 d { \bm  k}_2  .  
\eea
\end{widetext}
%
%
Equation (\ref{pa}) has an obvious exponential solution which corresponds to a zero flux
$F$:
$${\cal P}_k^{(1)} = {1 \over \langle J_k \rangle} \, e^{- s_k / \langle J_k \rangle}
$$
which corresponds to Gaussian statistics of the wave field $a_k$. However, there are also
solutions corresponding to $F=$const$\ne0$ which for $s_k \gg \langle J_k \rangle$ has a power-law
asymptotic \cite{cln1,cln2},
$$
{\cal P}_k^{(1)} = -{F  \over  \gamma_k s_k}.
$$
These solution corresponds to enhanced probability (with respect to gaussian) of strong
waves which is called intemittency of WT. Here, the constant flux in the amplitude space
$F$ can be associated with a wavebreaking process the exact form of which depends on the
physical system. For example, for the gravity water surface waves the wavebreaking process
takes form of whitecapping, and for the focusing NLS system the wavebreaking is represented
by filamentation/collapsing events. Obviously, this power-law tail of the PDF cannot extend to
infinity because the integral of the PDF must converge. Thus, there exists a cutoff which
can also be associated with the wave breaking, which can simply be understood that the
probability of waves with amplitude greater than some critical value must be zero.
Such critical value roughly corresponds to the amplitude for which the nonlinear term becomes
of the order of the nonlinear one so that the WT description breaks.

Multiplying equation (\ref{pa}) by $s_k^p$ and integrating over $s_k$, we have
the following equation for the moments
$ M^{(p)}_j = \langle J_j^{p} \rangle $:
\be
\frac{d }{d t}M^{(p)}_k = -p \gamma_k M^{(p)}_k +
p^2 \eta_k M^{(p-1)}_k.\label{moms}
\ee
which, for $p=1$ gives the kinetic equation for the waveaction spectrum,
\be
 \frac{d }{d t}n_k = - \gamma_k n_k + \eta_k .
\label{ke}
\ee
Substituting into this equation expressions for $\gamma_k$ and $\eta_k$, we obtain more
familiar forms of the kinetic equations:
\begin{widetext}
\bea
 \frac{d }{d t}n_k & =
& 4 \pi  \int
|V^k_{12}|^2
\delta(\o _k -\o _1-\o _2 )
\delta({ \bm  k} -  { \bm  k}_1 -{ \bm  k}_2)
  (n_{1} n_{2} - n_1 n_k - n_2 n_k)
\,  d { \bm  k}_1 d { \bm  k}_2   \nonumber \\
&&+
8 \pi  \int
|V^2_{k1} |^2
 \delta(\o _2 -\o _k-\o _1 )
\delta({ \bm  k}_2 -  { \bm  k} -{ \bm  k}_1)
(n_1n_2-n_1n_k+n_2n_k)
 \,  d { \bm  k}_1 d { \bm  k}_2  ,
\label{ke3}
\eea
and for the four-wave case
\BE\label{ke4}
 \frac{d }{d t}n_k=
4 \pi  \int
|T^{k1}_{23}|^2
\delta(\o _k +\o _1-\o _2 - \o _3) n_k n_1 n_2 n_3  \delta({ \bm  k}  + { \bm  k}_1 -{ \bm  k}_2-{ \bm  k}_3)   \left( {1 \over n_k} + {1 \over n_1} -
 {1 \over n_2}-{1 \over n_3} \right)
 \,  d { \bm  k}_1 d { \bm  k}_2  d {  \bm  k}_3\ .
\ee
\end{widetext}
Based on \Eqs{ke3} and (\ref{ke4}) [or \Eqs{pa} and (\ref{moms})] one can obtain the estimate
for the nonlinear frequency broadening in the WT regime, i.e. inverse characteristic
time of the nonlinear evolution as in (\ref{Ka}) or (\ref{Kb}).
 This leads to the WT applicability condition
(\ref{1Ka}) or (\ref{1Kb}).

Classical statistical  approach allows to obtain some interesting
and physically relevant solutions, such as Kolmogorov-Zakharov (KZ)
spectra corresponding to the energy and waveaction cascades through scales.
Such solutions can be obtained analytically using so-called Kraichnan-Zakharov transformation,
as well as from the scalings of the frequency and the interaction coefficients
based on the dimensional analysis. Discussion of these issues is beyond the scope of
our review, and the interested reader is referred for details to book
\cite{ZLF}. Here, it suffices to say that in most systems there exists
a shortcut way to obtain KZ spectra. It works for the systems with only one relevant
dimensional parameter, for example the gravity constant $g$ for the water surface
gravity waves, surface tension constant $\sigma$ for the capillary waves,
speed of sound $c_s$ for acoustic turbulence, quantum of circulation $\kappa$
for Kelvin waves on quantized vortex lines, etc. In this case the 1D energy spectrum
$E_k \sim k^\nu$ can be immediately obtained from the physical dimension of this  constant
which gives for the direct cascade \cite{conn}:
\be \label{kz_dir}
\nu = 2 \alpha +d-6 + \frac{5 - 3 \alpha -d}{N-1},
\ee
where $\alpha$ is the power of the dispersion relation $\omega \sim k^\alpha$
(which is uniquely determined by the above dimensional
constant), $d$ is the dimension of
the system and $N$ is the number of waves involved in the resonance
interaction.
For example, for the water surface gravity waves we have
$E_k \sim k^{-5/2}$, for the capillary waves $E_k \sim k^{-7/4}$ (both of these spectra are
called Zakharov-Filonenko spectra \cite{zakh_fil,fil}),
for acoustic turbulence $E_k \sim k^{-3/2}$ (Zakharov-Sagdeev spectrum \cite{ZakharovSagdeev}).
For Kelvin waves on quantized vortex lines, considering them as a local six-wave process,
one formally gets $E_k \sim k^{-7/5}$ (Kozik-Svistunov spectrum
\cite{kozik_svistunov}). However, this spectrum was recently shown in to be nonlocal

Similar approach one can use for finding the inverse cascade spectra, e.g. for the water surface
gravity waves or Kelvin waves \cite{conn}.

\end{document}